\documentstyle[aps]{revtex}

\draft

\begin{document}
\title{Generalized Ideal Gas Equations for Structureful Universe}
\author{Shahid N. Afridi$^{1}$ and M. Khalid Khan$^{2}$}
\address{Department of Physics, Quaid-i-Azam University,\\
Islamabad, Pakistan.\\
$^{1}$ email: snafridi@phys.qau.edu.pk\\
$^{2}$ email: mkk@qau.edu.pk}
\maketitle

\begin{abstract}
We have derived generalized ideal gas equations for a structureful universe
consisting of all forms of matters. We have assumed a universe that contains
superclusters. Superclusters are then made of clusters. Each cluster can be
further divided into smaller ones and so on. We have derived an expression
for the entropy of such a universe. Our model is rather independent of the
geometry of the intermediate clusters. Our calculations are valid for a
non-interacting universe within non-relativistic limits. We suggest that
structure formation can reduce the expansion rate of the universe.
\end{abstract}


\section{Introduction}

Ideal gas equations can be applied to non-interacting dilute gases.
Molecules and atoms in gas phase are so far away from each other, it makes a
little difference if we ignore the molecular interaction. These molecules
and atoms are further subject to Heisenberg uncertainty principle \cite{Lee}%
. When dealing with the universe as a whole, we see that the molecules,
nucleons and other subatomic particles are not uniformly distributed over
the space. There exists various scales in the universe. The nucleons are
particles in a star and at a larger scale the stars can be treated as
particles of a galaxy and so on. In order to incorporate all scales one is
required to generalize the ideal gas equation. In this paper, we have
derived generalized gas equations for such a universe.

In an earlier paper, based on configuration space, we addressed future
orientation of time with the growth of entropy \cite{Ak}. In the present
paper we have extended the same model to phase space. In this model we
assume that the universe is structureful which consists of clusters. Each
cluster contains subclusters. These subclusters can further be divided into
even smaller ones and so on. Such a description is called the universal
self-similarity \cite{Ak}. It should be noted that, for instance, cluster of
galaxies are bound, virialized, high over density system, held together by
the cluster self gravity \cite{Bahcall}. In this perspective we can say that
our model intrinsically incorporates gravity.

\section{Model}

We can assume that on large scale the universe is homogeneous, isotropic 
\cite{Wheeler} and unique \cite{Ak,CLO}. We can further assume that all
forms of the matter in the universe are contained in clusters. Each cluster
contains sub-clusters and so on. We treat these clusters, at their
corresponding scales, as particles of ideal gas. We confine ourself to
nonrelativistic regime.

We can represent the universe by a set $G^{n}$ \cite{Ak}

\[
G^{n}=\{G_{1}^{n{\small -}1},G_{2}^{n-1},....,G_{N_{n}}^{n-1}\}, 
\]
where $G_{i}^{n-1}$ represents the $i^{\text{th}}$ cluster and $N_{n}$ is
the number of clusters in the $G^{n}$. We can regard that $N_{n}$ is the
cardinal number of set $G^{n}$. Similarly we can write $G_{i}^{n-1}$ as 
\[
G_{i}^{n-1}=\{G_{1}^{n-2},G_{2}^{n-2},....,G_{N_{n-{\small 1}}}^{n-2}\}, 
\]
We can continue this to sub-atomic level and finally we get 
\[
G_{j}^{1}=\{G_{1}^{0},G_{2}^{0},....,G_{N_{{\small 1}}}^{0}\}, 
\]
where $G_{k}^{0}$ 's correspond to constituent particles which we assume,
have no further substructure.

The spatial distribution of accessible states for this system can be written
as \cite{Lee,Ak,Bowley} 
\begin{equation}
\Gamma _{n}^{\alpha }=\left( \frac{V_{n}^{\alpha }}{\Delta V_{n-1}^{\alpha }}%
\right) ^{N_{n}},  \eqnum{1}  \label{1}
\end{equation}
where $\alpha $ stands for configurations space. $V_{n}^{\alpha }$ is the
spatial volume of $G^{n}$ and $\Delta V_{n-1}^{\alpha }$ is the volume
occupied by $G_{i}^{n-1}$. We assume that $\Delta V_{n-1}^{\alpha }$ is not
arbitrarily small. It is much smaller as compared to $V_{n}^{\alpha }$ at
the scale of $G^{n}$ whereas at the scale of $G_{i}^{n-1}$, it is much
larger than $\Delta V_{n-2}^{\alpha }$. Therefore the corresponding
accessible states at the scale of $G_{i}^{n-1}$

\begin{equation}
\Gamma _{n-1}^{\alpha }=\left( \frac{\Delta V_{n-1}^{\alpha }}{\Delta
V_{n-2}^{\alpha }}\right) ^{N_{n-1}},  \eqnum{2}  \label{2}
\end{equation}
where $\Delta V_{n-2}^{\alpha }$ is the volume occupied by $G_{j}^{n-2}$.
Putting $\Delta V_{n-1}^{\alpha }$ from eq.(\ref{2}) into eq.(\ref{1}) and
on re-arranging we get 
\begin{equation}
\frac{V_{n}^{\alpha }}{\Delta V_{n-2}^{\alpha }}=\left( \Gamma _{n}^{\alpha
}\right) ^{1/N_{n}}\left( \Gamma _{n-1}^{\alpha }\right) ^{1/N_{n-1}} 
\eqnum{3}  \label{3}
\end{equation}
We can generalize it as follow

\begin{equation}
\frac{V_{n}^{\alpha }}{\Delta V_{0}^{\alpha }}=\prod\limits_{i=1}^{n}\left(
\Gamma _{i}^{\alpha }\right) ^{1/N_{i}}  \eqnum{4}  \label{4}
\end{equation}
As we know that 
\begin{equation}
s=k_{B}\ln \Gamma ,  \eqnum{5}  \label{5}
\end{equation}
where $s$ is the entropy and $k_{B}$ is the Boltzmann's constant. Using eq. (%
\ref{4}) and eq. (\ref{5}), we get 
\begin{equation}
\ln \left( \frac{V_{n}^{\alpha }}{\Delta V_{0}^{\alpha }}\right)
=\sum\limits_{i=1}^{n}\frac{s_{i}^{\alpha }}{k_{B}N_{i}}  \eqnum{6}
\label{6}
\end{equation}
where $s^{\alpha }$ is the spatial entropy and $\Delta V_{0}^{\alpha }$ is
the spread or uncertainty in volume of the constituent particles.

It is important to note that a relation describing average value of physical
quantities such as entropy, mass or thermal energy of an element of $G^{i}$
with an element of $G^{i-1}$ can be written as 
\begin{equation}
x_{i}=N_{i}x_{i-1}  \eqnum{7}  \label{7}
\end{equation}
where $x$ can be the average value of entropy, mass or thermal energy etc.
Using eq.(\ref{7}) for entropy into eq.(\ref{6}) and after iteration we get 
\begin{equation}
\ln \left( \frac{V_{n}^{\alpha }}{\Delta V_{0}^{\alpha }}\right) =\frac{%
{\cal N}s_{0}^{\alpha }}{k_{B}}  \eqnum{8}  \label{8}
\end{equation}
where $s_{0}^{\alpha }$ is the average of spatial entropy of the constituent
particle and 
\begin{equation}
{\cal N}=1+N_{1}+N_{1}N_{2}+....+\prod\limits_{i=1}^{n-1}N_{i}  \eqnum{9}
\label{9}
\end{equation}
It is worth mentioning the last term in the series of ${\cal N}$ is much
large as compared to other terms in the series.

It follows from eq. (\ref{7}) 
\begin{equation}
X\equiv x_{n}=\prod\limits_{i=1}^{n}N_{i}x_{0}=N_{t}x_{0}  \eqnum{10}
\label{10}
\end{equation}
where $X$ corresponds to the average value of entropy, mass, thermal energy
or any other additive physical quantity of the universe $G^{n}$. $N_{t}$
stands for total number of massive particles in the universe. In the above
equation, we have further assumed that each $G^{i}$ have equal number of $%
G^{i-1}$ elements.

If we treat $X$ as entropy in eq. (\ref{10}) and using this equation and eq.(%
\ref{8}), we can obtain the spatial entropy of the universe which can be
written as, 
\begin{equation}
S^{\alpha }=\frac{k_{B}N_{t}}{{\cal N}}\ln \left( \frac{V_{n}^{\alpha }}{%
\Delta V_{0}^{\alpha }}\right)  \eqnum{11}  \label{11}
\end{equation}

Next we consider momentum space. The momentum distribution of accessible
states for $G^{n}$ can be written as \cite{Lee}

\begin{equation}
\Gamma _{n}^{\beta }=\frac{V_{n}^{\beta }}{\Delta V_{n-1}^{\beta }} 
\eqnum{12}  \label{12}
\end{equation}
Here \cite{Lee,Web} 
\begin{equation}
V_{n}^{\beta }=\frac{\left( 2\pi m_{n-1}u_{n}\right) ^{3N_{n}/2}}{\left(
3N_{n}/2\right) !}  \eqnum{13}  \label{13}
\end{equation}
where superscript $\beta $ stands for momentum space. $m_{n-1}$ is the
average mass of $G_{i}^{n-1}$ and $u_{n}$ is the average kinetic energy of $%
G^{n}$.

Similarly we can write 
\begin{equation}
\Gamma _{n-1}^{\beta }=\frac{\Delta V_{n-1}^{\beta }}{\Delta V_{n-2}^{\beta }%
}  \eqnum{14}  \label{14}
\end{equation}
Following the same procedure as for eq. (\ref{4}), we can write 
\begin{equation}
\prod\limits_{i=1}^{n}\Gamma _{i}^{\beta }=\frac{V_{n}^{\beta }}{%
N_{1}!\Delta V_{0}^{\beta }},  \eqnum{15}  \label{15}
\end{equation}
where $\Delta V_{0}^{\beta }=\left( \Delta p_{x}\right) ^{3N_{1}}$and $%
\Delta p_{x}$ is the uncertainty in momentum of the constituent particle in
one dimension. As we reach smaller and smaller $G$'s, then we are ultimately
in quantum regime (i.e. nucleons in a star). A factor of $N_{1}!$ appears in
the denominator of r.h.s of eq. (\ref{15}) in order to avoid over counting
of $N_{1}$ momenta of identical nucleons in stars \cite{Lee}.

After doing some straight forward calculation we can write an expression for
entropy due to momentum distribution as

\begin{equation}
S^{\beta }=\frac{N_{t}k_{B}}{{\cal N}+N_{t}}\{\ln V_{n}^{\beta }-N_{1}\ln
N_{1}+N_{1}-3N_{1}\ln \Delta p_{x}\}  \eqnum{16}  \label{16}
\end{equation}
For large $N_{n}$ \cite{Web} 
\begin{equation}
\ln V_{n}^{\beta }\simeq \frac{3N_{n}}{2}\ln \left( \frac{2\pi e}{3N_{n}}%
.2m_{n-1}u_{n}\right)  \eqnum{17}  \label{17}
\end{equation}
where $e=2.718...$ i.e. base of natural logarithm. Using above approximation
we can rewrite eq. (\ref{16}) as 
\begin{equation}
S^{\beta }=\frac{N_{t}k_{B}}{{\cal N}+N_{t}}\left\{ \frac{3N_{n}}{2}\ln
\left( \frac{4\pi em_{n-1}u_{n}}{3N_{n}}\right) -N_{1}\ln
N_{1}+N_{1}-3N_{1}\ln \Delta p_{x}\right\}  \eqnum{18}  \label{18}
\end{equation}

Now using eqs. (\ref{11}) and (\ref{18}), we can write the combined entropy
of our universe in phase space as 
\begin{eqnarray}
S &=&S^{\alpha }+S^{\beta }  \eqnum{19}  \label{19} \\
&=&\frac{N_{t}k_{B}}{{\cal N}}\ln \left( \frac{V_{n}^{\alpha }}{\Delta
V_{0}^{\alpha }}\right) +\frac{N_{t}k_{B}}{{\cal N}+N_{t}}\left\{ \frac{%
3N_{n}}{2}\ln \left( \frac{4\pi em_{n-1}u_{n}}{3N_{n}}\right) -N_{1}\ln
N_{1}+N_{1}-3N_{1}\ln \Delta p_{x}\right\}  \nonumber
\end{eqnarray}
As from eqs. (\ref{7}) and (\ref{10}), we can write $m_{n-1}N_{n}=m_{n}%
\equiv M$ and $u_{n}\equiv U$, where $M$ is the mass of the universe, $U$
stands for the total thermal energy of the universe. We can finally write
entropy of the universe as 
\begin{equation}
S=\frac{N_{t}k_{B}}{{\cal N}}\ln \left( \frac{V}{\Delta V_{0}}\right) +\frac{%
N_{t}k_{B}}{{\cal N}+N_{t}}\left\{ \frac{3N_{n}}{2}\ln \left( \frac{4\pi eMU%
}{3N_{n}^{2}}\right) -N_{1}\ln N_{1}+N_{1}-3N_{1}\ln \Delta p_{x}\right\} , 
\eqnum{20}  \label{20}
\end{equation}
Here $N_{n}$ is the number of top most clusters (say the number of
superclusters in the universe) which makes the universe and $N_{1}$ is the
average number of particles in the bottom most cluster (say the number of
nucleons in a typical star). In the above equation we have also dropped the
indices $\alpha $ and $n$ over the spatial volume of the universe for
brevity.

\section{Implication of the Model}

We can now find the thermal energy and the equation of state by using the
following thermodynamics relations. 
\begin{equation}
T=\left( \frac{\partial U}{\partial S}\right) _{V,N}  \eqnum{21}  \label{21}
\end{equation}
and 
\begin{equation}
\frac{P}{T}=\left( \frac{\partial S}{\partial V}\right) _{U,N}  \eqnum{22}
\label{22}
\end{equation}
where $P$ is pressure and $T$ is temperature. From eqs. (\ref{20}) and (\ref
{21}), we get 
\begin{equation}
U=\frac{3N_{n}}{2}.\frac{N_{t}}{{\cal N}+N_{t}}k_{B}T  \eqnum{23}  \label{23}
\end{equation}
This equation gives us the thermal energy of our structureful universe. From
eqs. (\ref{20}) and (\ref{22}) 
\begin{equation}
PV=\frac{N_{t}k_{B}T}{{\cal N}},  \eqnum{24}  \label{24}
\end{equation}
Eqs. (\ref{23}) and (\ref{24}) can be treated as generalized equations for
an ideal gas for the structureful universe.

To verify validity of our system of equations, the ideal gas equations for
molecules/particles must be deduced from eqs. (\ref{23}) and (\ref{24}).
These equations can be obtained, if we take $n=1$ as a special case. In this
case $N_{n}=N_{t}$, where $N_{t}$ is the number of particles in a gas.
Further we find that for such a system ${\cal N}=1$, which can be neglected
as compared to $N_{t}$ in the denominator of r.h.s. of eq. (\ref{23}). We
finally get 
\begin{equation}
U=\frac{3}{2}N_{t}k_{B}T  \eqnum{25}  \label{25}
\end{equation}
and 
\begin{equation}
PV=N_{t}k_{B}T  \eqnum{26}  \label{26}
\end{equation}
The last two equations give us the thermal energy and the equation of state
for an ideal gas respectively.

We can get interesting results from these equations. Let us denote the
thermal energy (given in eq. (\ref{23}) of the structureful universe by $%
U^{A}$ and the pressure by $P^{A}$ (given in eq. (\ref{24}), and denoting
the thermal energy and pressure for structureless universe (given in eqs. (%
\ref{25}) and (\ref{26}) by $U^{B}$ and $P^{B}$ respectively. For same
temperature and volume 
\begin{equation}
\frac{U^{A}}{U^{B}}=\frac{N_{n}}{{\cal N}+N_{T}}  \eqnum{27}  \label{27}
\end{equation}
\begin{equation}
\frac{P^{A}}{P^{B}}=\frac{1}{{\cal N}}  \eqnum{28}  \label{28}
\end{equation}
As we can see that $U^{A}\ll U^{B}$, and $P^{A}\ll P^{B}$. It suggests that
expansion of structureful universe can be much slower as compared to a
structureless universe.

\section{Conclusion}

We have considered different scales in expansion of universe starting from
sub-atomic scale to super-clusters and beyond. We have obtained generalized
ideal gas equations that are applicable to our structureful dynamical
universe. It can be applied to stars, galaxies, cluster, supercluster and of
course to the universe as a whole. Our model is rather independent of the
shape and size of the intermediate clusters. In these derivations we were
confined to a non-interacting and non-relativistic system. Using these
results and astrophysical data one can quantitatively compute entropy and
other thermodynamical observables of the structureful universe.

\section{Acknowledgments}

We have benefited from useful comments by Ariel Caticha. One of us (SNA)
would like to acknowledge the ICSC- World Laboratory for providing partial
financial support.

\end{document}